\newcommand{\tstern}{\ensuremath{T_2^*} }
\def \deg {\ensuremath{^\circ} }

\newcommand*{\fg}[1]{Fig.\thinspace\ref{#1}}

\documentclass[aps,prl,preprint,superscriptaddress]{revtex4}

\usepackage{graphicx}
\usepackage{dcolumn}

\usepackage{bm}
\usepackage[english]{babel}

\begin{document}

%

\title{Electric field-driven coherent spin
reorientation of optically generated electron spin packets in InGaAs}

%

\author{S. Kuhlen}
\author{K. Schmalbuch}
\author{M. Hagedorn}
\author{P. Schlammes}
\author{M. Patt}\affiliation{II. Physikalisches Institut, RWTH Aachen University, 52056 Aachen, Germany}\affiliation{JARA-Fundamentals of Future Information Technology, J\"{u}lich-Aachen Research Alliance, Germany}
\author{M. Lepsa}\affiliation{Peter Gr\"{u}nberg Institut (PGI-9), Forschungszentrum J\"{u}lich GmbH, D-52425 J\"{u}lich, Germany}
  \affiliation{JARA-Fundamentals of Future Information Technology, J\"{u}lich-Aachen Research Alliance, Germany}
\author{G. G\"{u}ntherodt}
\author{B. Beschoten}
  \thanks{ e-mail: bernd.beschoten@physik.rwth-aachen.de}
  \affiliation{II. Physikalisches Institut, RWTH Aachen University, 52056 Aachen, Germany}
  \affiliation{JARA-Fundamentals of Future Information Technology, J\"{u}lich-Aachen Research Alliance, Germany}

\date{\today}

%

\begin{abstract}

Full electric-field control of spin orientations is one of the key
tasks in semiconductor spintronics. We demonstrate that electric
field pulses can be utilized for phase-coherent $\pm\pi$ spin
rotation of optically generated electron spin packets in InGaAs
epilayers detected by time-resolved Faraday rotation. Through
spin-orbit interaction, the electric-field pulses act as local
magnetic field pulses (LMFP). By the temporal control of the LMFP,
we can turn on and off electron spin precession and thereby rotate
the spin direction into arbitrary orientations in a 2-dimensional
plane. Furthermore, we demonstrate a spin echo-type spin drift
experiment and find an unexpected partial spin rephasing, which is
evident by a doubling of the spin dephasing time.

\end{abstract}

\maketitle
%

Most device concepts in semiconductor spintronics rely on the
efficient generation of spin-polarized carriers and their phase
sensitive manipulation and read-out. Initial experiments comprised
ferromagnet/semiconductor hybrid structures, where the ferromagnet
is either used as a source of spin-polarized carriers \cite{
Nature402_Ohno1999_ElectricalSpinInjectioninaFerromagneticSemiconductorHeterostructure,
Science309_Crooker2005_ImagingSpinTransportinLateralFerromagnet-SemiconductorStructures,NatureMaterials3_Kioseoglou2004_ElectricalSpinInjectionfromann-typeFerromagneticSemiconductorintoaIII-VDeviceHeterostructure}
or as a spin-sensitive detector using magneto-resistive read-out
\cite{LOU2007_electrical_detection}. In recent years, however, a new
pathway towards spintronics without ferromagnets has evolved, which
allows to generate and to manipulate spins by electric fields $E$
only \cite{Awschalom_Trends,kato2003}. In ordinary non-magnetic
semiconductors, dc $E$ fields can generate spins by two
complementary effects, the spin Hall effect \cite{Hirsch_PRL,
Zhang_PRL,
Science306_Kato_Observation_of_the_Spin_Hall_Effect_in_Semiconductors,
PhysRevLett.94.047204, Sih_Nat_Phy_Spatial_imaging,
PhysRevLett.105.156602} and the so-called current induced spin
polarization (CISP)
\cite{SSC73_Edelstein1990_SpinPolarizationofConductionElectronsInducedbyElectricCurrentinTwo-DimensionalAsymmetricElectronSystems,
PRL93_Kato2004_Current-InducedSpinPolarizationinStrainedSemiconductors,
PhysRevLett.97.126603, Koehl_CISP_GaN}. Both result from the
spin-orbit (SO) coupling. The spin Hall effect leads to a spin
accumulation transverse to the current flow direction by spin
dependent scattering
\cite{Science306_Kato_Observation_of_the_Spin_Hall_Effect_in_Semiconductors},
while CISP is manifested by a uniform spin polarization in the
semiconductor, which has been demonstrated both by static and by
time-resolved magneto-optical probes
\cite{PRL93_Kato2004_Current-InducedSpinPolarizationinStrainedSemiconductors}.
Although the microscopic origin of CISP is not fully understood
\cite{SovJETPL50_Aronov1989_NuclearElectricResonanceandOrientationofCarrierSpinsbyanElectricField,SSC73_Edelstein1990_SpinPolarizationofConductionElectronsInducedbyElectricCurrentinTwo-DimensionalAsymmetricElectronSystems,PRL98_Engel2007_Out-of-PlaneSpinPolarizationfromIn-PlaneElectricandMagneticFields,PRB78_Liu2008_Current-InducedSpinPolarizationinSpin-Orbit-CoupledElectronSystems},
in most systems electron spins get oriented along the effective
internal magnetic field $B_{int}$, which can be tuned by the $E$
field strength through SO coupling. Internal magnetic fields have
also been determined in 2D electron gases by Shubnikov-de Haas
oscillations
\cite{PRB39_Das1989_EvidenceforSpinSplittinginIn_xGa_1-xAs-In_0-52Al_0-48AsHeterostructuresasB-0,
PRB55_Engels1997_ExperimentalandTheoreticalApproachtoSpinSplittinginModulation-DopedIn_xGa_1-xAsInPquantumwellsforB-0},
antilocalization
\cite{PRL89_Koga2002_RashbaSOCouplingProbedbytheWeakAntilocalizationAnalysisinInAlAs-InGaAs-InAlAsQWasaFunctionofQWAsymmetry},
photo-current
\cite{PRL92_Ganichev2004_ExperimentalSeparationofRashbaandDresselhausSpinSplittingsinSemiconductorQuantumWells},
static Hanle \cite{kalevich1990,crooker2007} and time-resolved
Faraday rotation (TRFR) measurements
\cite{NaturePhysics3_Meier2007_MeasurementofRashbaandDresselhausSpin-OrbitMagneticFields,
PRB82_Norman2010_MappingSpin-OrbitSplittinginStrained(InGa)AsEpilayers}.
The control of $B_{int}$ is of fundamental importance for spin
manipulation. It can be realized by gate voltages in 2D electron
gases
\cite{PRL78_Nitta1997_GateControlofSpin-OrbitInteractioninanInvertedInGaAs-InAlAsHeterostructure}
or by dc $E$ fields
\cite{Gosh_electric_control_of_pin_coherence_in_ZnO,
Nature427_Kato2004_CoherentSpinManipulationwithoutMagneticFieldsinStrainedSemiconductors}.
These SO fields have been used by Kato \emph{et al.} to induce spin
precession at zero external magnetic field $B_{ext}$
\cite{Nature427_Kato2004_CoherentSpinManipulationwithoutMagneticFieldsinStrainedSemiconductors}.

In this Letter we report on TR electrical spin manipulation
experiments of electron spins in InGaAs. Coherent spin packets are
optically generated by circularly polarized laser pump pulses. Their
initial spin direction is manipulated by $E$ field pulses, which act
as effective local magnetic field pulses due to SO coupling. Using
TRFR we probe the Larmor precession of spin packets induced by the
SO field pulse. By changing the pulse width and polarity we are able
to rotate the spins into arbitrary directions within a 2-dimensional
plane. In addition to spin precession, the $E$ field pulses also
yield a lateral drift of the spin packet over several $\mu$m. As
sign reversal of the pulses will reverse both spin precession and
drift direction, we are able to explore spin echo of the spin packet
in the diffusive spin transport regime.

\begin{figure}[tbp]
\includegraphics{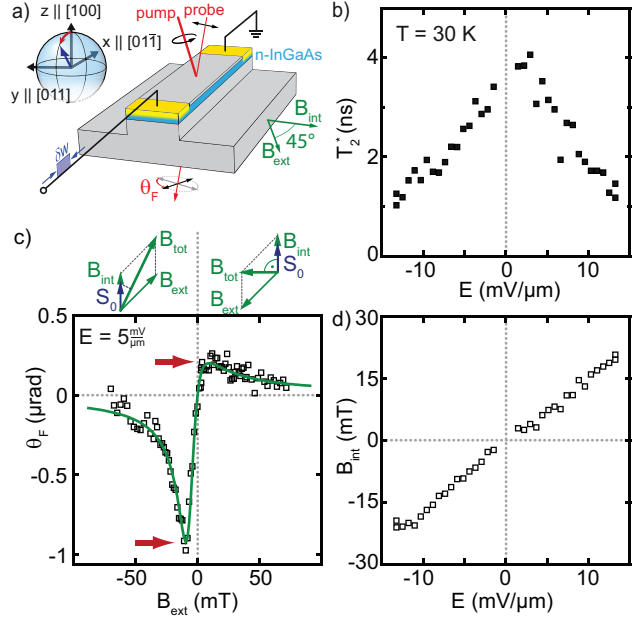}
\caption{ \label{fig1} (color online). (a)~Schematic setup. DC or
pulsed $E$ fields are applied along the $[01\bar1]$ direction of an
$n$-InGaAs transport channel. Electrically or optically generated
spins are probed by static and TRFR in polar geometry. (b)~\tstern
vs. dc $E$ field. (c)~(upper panel) Schematics of total effective
magnetic field $\mathbf{B}_{tot}$ for positive and negative
$B_{ext}$. (lower panel) Asymmetric Hanle curve taken at $E=+
5\,\frac{\mathrm{mV}}{\mathrm{\mu m}}$ and $T=30$~K. The green line
is a fit to Eq.~1. (d)~$B_{\mathit{int}}$ determined from asymmetric
Hanle signal vs. dc $E$ field. }
\end{figure}

Our studies were performed on a 500~nm thick
In$_{0.07}$Ga$_{0.93}$As epilayer grown on semi-insulating (100)
GaAs by molecular beam epitaxy. The room temperature carrier density
was set to \mbox{$n\sim 3\times 10^{16}$~cm$^{-3}$} by Si-doping to
allow for long spin dephasing times at low temperatures
\cite{PRB66_Dzhioev2002_Low-temperatureSpinRelaxationinN-typeGaAs,PRL80_Kikkawa1998_ResonantSpinAmplificationinN-TypeGaAs}.
By chemical wet etching a 140~$\mu$m wide and 680~$\mu$m long
transport channel was patterned and contacted with standard Au/Ge/Ni
electrodes. For spin manipulation experiments, the electric field
was applied along the $[01\bar1]$ crystal axis (or x axis) as shown
in \fg{fig1}a. For our samples, this configuration yields the
strongest CISP with internal magnetic fields pointing along the
$[011]$ or $y$ axis (see also
\cite{PRL93_Kato2004_Current-InducedSpinPolarizationinStrainedSemiconductors}).
The device is embedded in a coplanar wave guide and connected to
microwave probes \cite{supplementary}. In the following, we discuss
two classes of experiments: (I) In static CISP, we will use dc $E$
fields to probe the $E$ field induced spin polarization measuring
the Faraday rotation $\theta_F$ in polar geometry (along the $[100]$
or $z$ axis). From the shape of the resulting Hanle depolarization
curves we are able to directly extract the internal magnetic field
strength $\mathbf{B}_{\mathit{int}}$ at non-collinear alignment of
$\mathbf{B}_{\mathit{int}}$ with the external magnetic field
$\mathbf{B}_{\mathit{ext}}$ (see \fg{fig1}a). (II) In TRFR
experiments, we use circularly polarized ps laser pump pulses to
trigger electron spin coherence in InGaAs. Spin precession about the
vector sum of $\mathbf{B}_{\mathit{int}}$ and
$\mathbf{B}_{\mathit{ext}}$ is probed by a second time-delayed
linearly polarized probe pulse using TRFR measurements.
$\mathbf{B}_{\mathit{int}}$ can either result from dc or pulsed
electric fields. The latter stems from a pulse-pattern generator,
which is synchronized to the picosecond pump laser and is used for
time-resolved spin reorientation and spin echo-type experiments
\cite{supplementary}.

We first use dc-CISP to determine the direction of
$\mathbf{B}_{\mathit{int}}$ and its magnitude in our InGaAs
structures. With $\mathbf{B}_{int}$ being not perpendicular to
$\mathbf{B}_{ext}$ ($\alpha=\angle(\mathbf{B}_{ext},
\mathbf{B}_{int})\neq 90\deg$) the symmetry of the otherwise
expected anti-symmetric Hanle curve is broken. For example in
\fg{fig1}c the amplitude of $|\theta_F|$ at its extremal values
varies by a factor of three (see red arrows), which can be
attributed to the influence of $\mathbf{B}_{\mathit{int}}$ on the
precession axis and frequency as the spins precess about
$\mathbf{B}_{tot}$ \cite{crooker2007}. As illustrated at the top of
\fg{fig1}c, the respective magnitudes of $\mathbf{B}_{tot}$ differs
for $\pm\mathbf{B}_{ext}$ and the angle between $\mathbf{B}_{tot}$
and the initial spin orientation $\mathbf{S}_0$ changes
significantly.

Assuming $\mathbf{B}_{\mathit{int}}\perp z$ and
$\mathbf{S}_0\parallel\mathbf{B}_{\mathit{int}}$, we can model
\cite{supplementary} the Hanle curves by
\begin{equation}
 \theta_F(B_{\mathit{ext}})=\theta_0\cdot{\frac{B_{\mathit{ext}}\cdot\sin \alpha}{B_{1/2}}}\cdot\left[{1+\left( \frac{B_{\mathit{tot}}}{B_{1/2}}\right)^2}\right]^{-1},
\label{Fitformel}
\end{equation}

with amplitude $\theta_0\propto S_0$, the total effective magnetic
field
$\mathbf{B}_{\mathit{tot}}=\mathbf{B}_{\mathit{int}}+\mathbf{B}_{\mathit{ext}}$
and the angle $\alpha$ between $\mathbf{B}_{\mathit{ext}}$ and
$\mathbf{B}_{\mathit{int}}$ (i.e. $45\deg$). The width of the Hanle
curve $B_{1/2}$ is a direct measure of the transverse dephasing time
$\tstern=(g\frac{\mu_B}{\hbar} B_{1/2})^{-1}$.

The Hanle curves can be fitted
according to Eq.~1 (see green curve in \fg{fig1}c). As seen in \fg{fig1}b, we observe a strong
decrease of $\tstern$ for both $E$ field polarities, which has also
been observed in Ref. \cite{PRL93_Kato2004_Current-InducedSpinPolarizationinStrainedSemiconductors}
indicating additional $E$ field dependent spin dephasing. The extracted
$B_{int}$ values in \fg{fig1}d vary almost linearly with the $E$ field
and vanish at $E=0$ (see also
\cite{NaturePhysics3_Meier2007_MeasurementofRashbaandDresselhausSpin-OrbitMagneticFields}). These internal magnetic fields will be used next for coherent spin manipulation.

\begin{figure}[bp]
\includegraphics{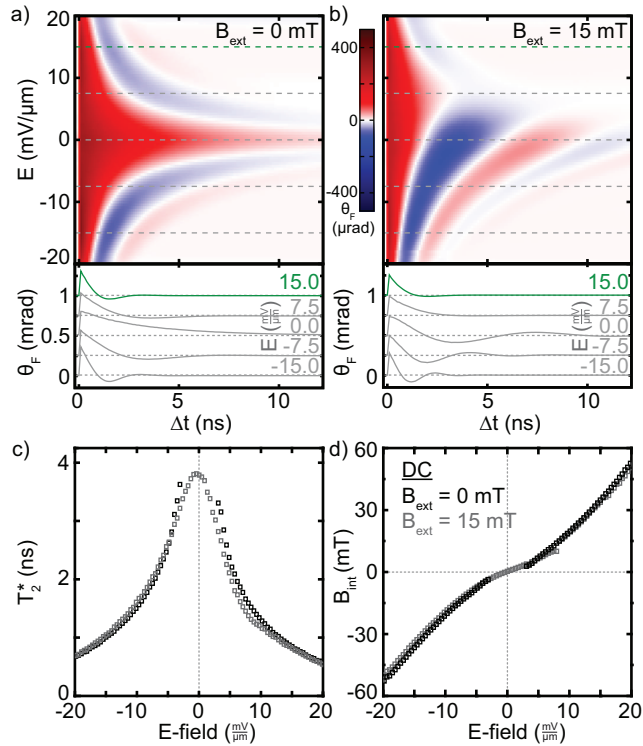}
\caption{ \label{fig2} (color online). TRFR after optical spin
orientation in InGaAs ($T=30$~K). Electron drift in an  $E$ field
induces $B_{int}$ which results in spin precession about
$\mathbf{B}_{\mathit{tot}}=\mathbf{B}_{\mathit{int}}+\mathbf{B}_{\mathit{ext}}$,
shown for (a)~$B_{\mathit{ext}}=0$\,mT and
(b)~$B_{\mathit{ext}}=15$\,mT. TRFR scans are
plotted in the lower parts at selected $E$ fields. An offset is
added for clarity. The resulting parameters (c)~$\tstern$ and
(d)~$B_{\mathit{int}}$ are in good agreement with the values
extracted from Hanle measurements (\fg{fig1}b and \fg{fig1}d). }
\end{figure}

For this purpose, coherent electron spin ensembles are generated
along the z direction by circularly polarized picosecond laser pump
pulses
\cite{PRL80_Kikkawa1998_ResonantSpinAmplificationinN-TypeGaAs,
PhysRevLett.105.246603} and detected by TRFR in polar geometry. The
$E$ field will now be used for TR spin manipulation. We note that
the $E$ field pulse itself can also create a phase triggered
coherent spin polarization, which can be probed by TR-CISP
\cite{PRL93_Kato2004_Current-InducedSpinPolarizationinStrainedSemiconductors}.
This effect is, however, negligible as the fraction of spin
polarization by CISP is three orders of magnitude less than the spin
polarization obtained after optical orientation.

We first explore the influence of dc $E$ fields on the coherent spin
ensemble in Figs.~2a and 2b at $B_{ext}=0$ and 15~mT, respectively.
From these experiments it is obvious that SO induced electron spin
precession can be triggered by electrical means. In the former case,
for both negative and positive $E$ fields of the same magnitude the
spins precess with equal Larmor frequencies $\omega_L=g
\frac{\mu_B}{\hbar}B_{\mathit{tot}}$, where $g$ is the electron
g-factor, $\mu_B$ Bohr's magneton, and $\hbar$ Planck's constant. In
contrast, in the latter case (see \fg{fig2}b) spin precession is
accelerated for $E<0\,\frac{\mathrm{mV}}{\mathrm{\mu m}}$ while it
is slowed down for $0<E<7.5\,\frac{\mathrm{mV}}{\mathrm{\mu m}}$.
This dependence proves the reversal of the
$\mathbf{B}_{\mathit{int}}$ direction upon sign reversal of
$\mathbf{E}$. For $E=0\,\frac{\mathrm{mV}}{\mathrm{\mu m}}$ multiple
spin precessions can be observed due to the enhanced $\tstern$.

All TRFR data can be described by an exponentially damped cosine
function
\begin{equation}
\theta_F(\Delta t)=\theta_0\cdot\exp\left(-\frac{\Delta t}{\tstern}\right)\cdot\cos\left(\omega_L \Delta t+\delta\right),
\label{gedOszillation}
\end{equation}
with amplitude $\theta_0$, pump-probe delay $\Delta t$ and phase
$\delta$. This way we can determine $\tstern$ and
$B_{\mathit{int}}$, which are plotted vs $E$ in Figs. 2c and 2d,
respectively. We note a strong decrease of $\tstern$, which limits
the observable spin coherence. The decrease of $\tstern$ has
previously been assigned to spins drifting out of the probe laser
focus
\cite{Nature427_Kato2004_CoherentSpinManipulationwithoutMagneticFieldsinStrainedSemiconductors}.
However, as the dephasing times extracted from the above CISP
measurements (cp. to \fg{fig1}b) exhibit a similar decrease with $E$
field, we attribute this effect to additional spin dephasing, which
will be evaluated in more detail below by using $E$ pulses for spin
manipulation. $B_{int}$ values from TRFR depend also nearly linearly
on the $E$ field (Fig.~2d). The non-linear behavior around $E=0$
results from non-ohmic contact resistance as seen by the linear
dependence of $B_{int}$ on $I$ (not shown).

\begin{figure}[tbp]
\includegraphics{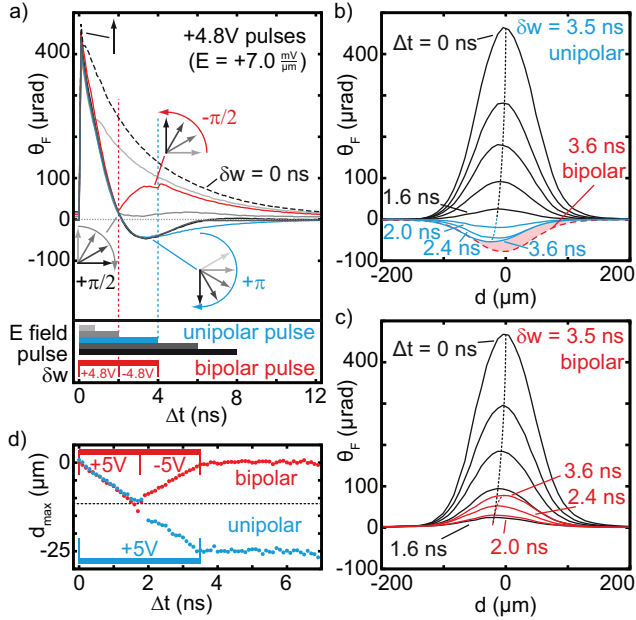}
\caption{ \label{fig3} (color online). Spin manipulation by unipolar
(blue) and bipolar (red) $E$ field pulses. All $E$ field pulses
start at $\Delta t=0$~ns. (a)~TRFR measurements of optically created
spin packets ($T=30$~K), which precess at $B_{ext}=0$ during $E$
field pulses of different width $\delta w$, which is visualized in
the lower panel. Spatio-temporal evolution of spin packet after spin
manipulation by (b) unipolar and (c) bipolar pulse. The temporal
spacing is $0.4$~ns. The final spin distribution at $\Delta t =
3.6$~ns in (c) has been reversed and added as a dashed line in (b).
(d)~Spatio-temporal drift of spin packet after spin manipulation
with unipolar (blue) and bipolar (red) pulses of equal pulse width
and magnitude.}
\end{figure}

\begin{figure}[tbp]
\includegraphics{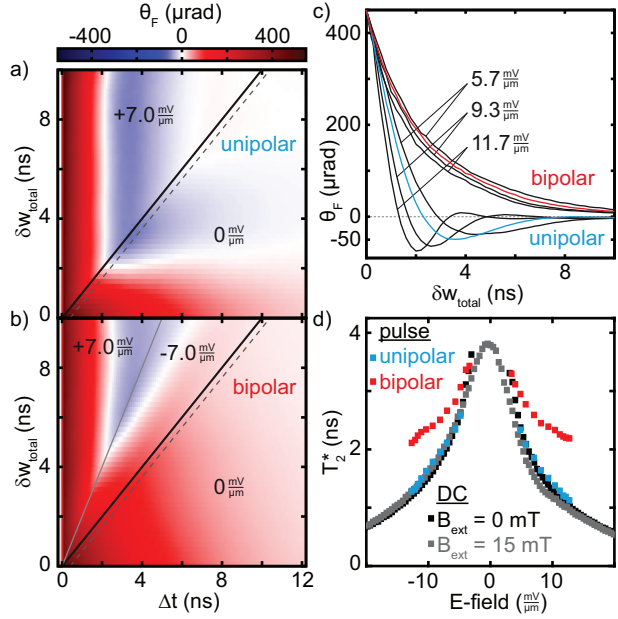}
\caption{ \label{fig4} (color online). False color plots of TRFR
measurements for spin manipulation by (a) unipolar and (b) bipolar
$E$ field pulse with variable pulse width $\delta w$. The end of the
pulses is marked by solid lines. (c)~$\theta_F(\Delta t = \delta w +
200$~ps) vs. $\delta w$ from traces along dashed lines in (a) and
(b). (d)~\tstern taken from fits to curves in (c). \tstern values
during unipolar pulses agree with results from dc $E$ fields (cp. to
\fg{fig2}c), while \tstern after bipolar pulses shows partial
rephasing.}
\end{figure}

While in the above dc experiments we can control the spin precession
frequency by $E$ fields only, we now want to manipulate the phase of
the optically generated coherent spin packet. In other words, we
will use $E$ field pulses both to initialize and to stop spin
precession at $B_{ext}=0$~mT. When the $E$ field pulse reaches the
optically generated spin packet, it will create a LMFP for the
duration of the pulse. This LMFP will trigger spin precession in the
$zx$-plane (see \fg{fig1}a). The precession frequency depends on the
$E$ field strength, while the total precession time is given by the
pulse width $\delta w$. In \fg{fig3}a we show a sequence of TRFR
measurements of optically generated coherent spin packets, which are
manipulated by $E$ field pulses of
$E=7~\frac{\mathrm{mV}}{\mathrm{\mu m}}$ and various pulse widths
ranging from 0 to 8~ns. As expected, we observe spin precession for
long pulses of 8~ns (black curve). As the laser repetition time is
12~ns, this case is close to the dc Hanle limit (\fg{fig2}a). For
shorter pulse widths, $\theta_F$ always follows this spin precession
curve during the field pulse. However, spin precession abruptly
stops after the pulse has turned off with elapsed $\delta w$. This
is seen by a simple exponential decay thereafter, which is observed
for all pulse widths. The minimum in $\theta_F$ at 3.7~ns shows that
the LMFP of 4~ns operates as $\pi$ pulse (blue curve in \fg{fig3}a),
which rotates the spins by $180\deg$ from the $+z$ into the $-z$
direction \cite{footnote}. The difference between a decaying signal
after a 4~ns pulse and further precession (i.e. for 6~ns and 8~ns
pulses) is clearly visible.

As discussed in \fg{fig2}d, reversing the $E$ field polarity will
reverse $B_{int}$, which results in a reversal of spin precession
direction. When using a bipolar pulse sequence which consists of two
subsequent pulses with opposite polarity and equal width and
magnitude, we expect spin reorientation of the spin packet to its
original direction at the end. The red curve in \fg{fig3}a shows
spin manipulation by a bipolar pulse sequence with the same
magnitude (4.8~V) and total width (4~ns) as the blue curve for spin
manipulation by a unipolar pulse. During the first 2~ns both
unipolar and bipolar pulses rotate the spin packet by $\pi/2$ into
the sample plane. While spin precession for the unipolar pulse will
continue to $\pi$ rotation, spin precession is reversed during the
subsequent 2~ns for bipolar pulses, which function as $-\pi/2$
pulses. Remarkably, at 4~ns the value $|\theta_F|$ is larger for the
bipolar pulse than for the unipolar pulse but less than the value
obtained for free decay of the ensemble (see black curve for $\delta
w=0$). As $|\theta_F|$ is a direct measure of the net spin moment,
its increase indicates partial spin rephasing.

To exclude that the different amplitudes result from $E$ field
induced drift of the spin packet away from the probe laser spot, we
show a series of spatio-temporal profiles of the spin packet in
Figs.~3b and 3c after unipolar and bipolar spin manipulation,
respectively. The data have been taken by scanning the probe
relative to the pump beam along the $E$ field direction. While the
spin packet drifts continuously to the left for unipolar pulses, it
reverses the drift direction after 2~ns for bipolar pulses and
returns to its original position thereafter (see also Fig. 3d). To
better compare the final spin distributions at 3.6~ns for both
manipulation schemes, we added the respective curve from Fig.~3c
with reversed sign as a red dashed line in Fig.~3b. Despite their
small difference in peak positions the dashed red curve has an
overall larger magnitude showing that drift effects are too small to
account for the difference in amplitudes and that the observed
effect is indeed caused by rephasing.

In the following we extract spin dephasing times for both spin
manipulation experiments to further quantify the effect of spin
rephasing. For each $E$ field value we have measured TRFR for pulse
widths ranging from 200~ps to 10~ns. In Figs.~4a and b we show the
respective $\theta_F$ vs $\Delta t$ curves on false color plots for
unipolar and bipolar pulses with
$|E|=7~\frac{\mathrm{mV}}{\mathrm{\mu m}}$. The solid black lines
mark the end of the pulses. The resulting $\theta_F$ after spin
manipulation is plotted vs $\delta w_{total}$ in Fig.~4c (blue curve
for unipolar pulses and red curve for bipolar pulses). These data
are taken at 200~ps after the end of each pulse (see dotted lines in
Figs.~4a and b). Spin precession can be observed for unipolar pulses
for pulse widths above 2~ns (see also Fig.~3). In contrast, no spin
precession or sign reversal of $\theta_F$ is seen for bipolar
pulses. Instead, $\theta_F$ is exponentially decaying unambiguously
demonstrating that the spin packet points along the original
direction after the bipolar pulse sequence. We also included in
Fig.~4c additional TRFR traces at selected $E$ field values which
have been extracted by the same method. It is obvious that
$\theta_F$ from spins precessing continuously in one direction
(unipolar pulse) decays much faster than the signal stemming from
bipolar pulses. This is most clearly seen for large $E$ field
values. The extracted spin dephasing times are depicted in Fig.~4d.
As expected, spin dephasing during the unipolar pulses (blue
squares) matches values from the above dc case (black squares). In
contrast, spin dephasing times after bipolar spin manipulation are
longer at all $E$ fields demonstrating that the bipolar pulse
sequence allows for spin-echo studies of the spin ensemble in
diffusive transport. The observed spin rephasing is strongest at the
largest $E$ fields where the spin dephasing times doubles.

Our findings show that the $E$ field induced decrease of $\tstern$
is partly caused by spin dephasing and not by spin relaxation.
During Larmor precession, a phase spreading is built up when the
LMFP is applied, which might result from local fluctuation of
$B_{int}$ across the spin packet. In contrast to standard spin echo
techniques we reverse the precession direction by changing the LMFP
polarity. As the spin ensemble now precesses in the opposite
direction, it can partially compensate for the accumulated phase
spreading. However, this is only true if the variation in precession
frequencies is identical for both drift directions for the
individual spins. We note that the observed spin rephasing is not
expected for Elliott-Yafet spin scattering \cite{Elliott(1954)} as
spin-flip events which occur during momentum scattering will destroy
time-reversal symmetry. In contrast, spin scattering due to
D'yakonov-Perel' mechanism \cite{DP(1954)} occurs between momentum
scattering events by spin precession about $k$-dependent spin-orbit
field. As momentum scattering occurs on ps time-scales
\cite{footnote1}, which is shorter than $T_2^*$ by 3 orders of
magnitude, the ensemble phase will be randomized during transport
for each momentum scattering event. As individual electrons will not
follow their identical paths during the spin echo pulse they will
precess about different spin-orbit fields, which also should not
result in spin rephasing in diffusive transport.

In conclusion, we have shown to achieve full time-resolved
electrical phase control of electron spin packet orientation within
a 2D plane in InGaAs in zero magnetic field. A novel spin-echo
technique has been used to explore electric field-induced spin
dephasing, which surprisingly revealed that partial rephasing is
possible even in diffusive spin transport. Some of us have shown
recently that linearly polarized light can be utilized to achieve
full 2D control of the initial spin direction
\cite{PhysRevLett.105.246603}. Adding the electric field driven spin
rotation to this new technique, we expect being able to achieve full
3D control of the spin orientation \cite{footnote2}, which could
provide an important step toward all-electrical spintronics without
ferromagnets.

%

We acknowledge useful discussion with F. Hassler. This work was
supported by DFG through FOR 912.

\bibliographystyle{apsrev}

\selectlanguage{english}

\begin{thebibliography}{28}
\bibitem{footnote}
Although $\pi$ rotation is completed at 4~ns, the minimum of
$\Theta_F$ is seen at 3.7~ns. This shift results from a strong
amplitude drop by the short $T_2^*$.
\bibitem{footnote1}
From the sample mobility of 5360 $\text{cm}^2/\text{Vs}$ at $T=30$~K
we extract a momentum relaxation time less than 1~ps which is
significantly shorter than the ns time-scale for spin manipulation.
\end{thebibliography}




\end{document}